\documentclass[sigconf,nonacm]{acmart}
\AtBeginDocument{%
  \providecommand\BibTeX{{%
    \normalfont B\kern-0.5em{\scshape i\kern-0.25em b}\kern-0.8em\TeX}}}

\usepackage{amsmath,amsfonts}
\usepackage{algorithmic}
\usepackage{graphicx}
\usepackage{textcomp}
\usepackage{xcolor}
\usepackage{caption}
\usepackage{subcaption}
\usepackage{booktabs}
\usepackage{comment}
\usepackage{cancel}
\usepackage{multicol}
\usepackage[moderate]{savetrees} 
\usepackage{numprint}

\usepackage[normalem]{ulem}

\newcommand{\ie}{\textit{i}.\textit{e}., }

\ifdefined\draftversion

    \newcommand{\frar}[2]{{\textcolor{gray}{\sout{#1}}\ }{\textcolor{green}{#2}}}
\else

    \newcommand{\frar}[2]{#2}
\fi

\copyrightyear{2024}
\acmYear{2024}
\setcopyright{rightsretained}
\acmConference[DAC '24]{61st ACM/IEEE Design Automation Conference}{June 23--27, 2024}{San Francisco, CA, USA}
\acmBooktitle{61st ACM/IEEE Design Automation Conference (DAC '24), June 23--27, 2024, San Francisco, CA, USA}
\acmDOI{10.1145/3649329.3657319}
\acmISBN{979-8-4007-0601-1/24/06}

\begin{document}

\title[Deep Reinforcement Learning based Online Scheduling Policy for DNN Multi-Tenant Multi-Accelerator Systems]{Deep Reinforcement Learning based Online Scheduling Policy for Deep Neural Network Multi-Tenant Multi-Accelerator Systems}

\author{Francesco G. Blanco, Enrico Russo, Maurizio Palesi, Davide Patti, Giuseppe Ascia, Vincenzo Catania}
\authornote{F. G. Blanco and E. Russo contributed equally to this research.}
\affiliation{%
  \institution{blanco.francesco@studium.unict.it, enrico.russo@phd.unict.it, \{name.surname\}@unict.it}
  \institution{University of Catania}
  \city{Catania}
  \country{Italy}
}


\begin{abstract}
\ifdefined\longversion
Deep Learning, particularly Deep Neural Networks (DNNs), has emerged as a powerful tool for addressing intricate real-world challenges. Nonetheless, the deployment of DNNs presents its own set of obstacles, chiefly stemming from substantial hardware demands. In response to this challenge, Domain-Specific Accelerators (DSAs) have gained prominence as a means of executing DNNs, especially within cloud service providers offering DNN execution as a service.
\fi

Currently, there is a growing trend of outsourcing the execution of DNNs to cloud services.
For service providers, managing multi-tenancy and ensuring high-quality service delivery, particularly in meeting stringent execution time constraints, assumes paramount importance, all while endeavoring to maintain cost-effectiveness. In this context, the utilization of heterogeneous multi-accelerator systems becomes increasingly relevant. This paper presents RELMAS, a low-overhead deep reinforcement learning algorithm designed for the online scheduling of DNNs in multi-tenant environments, taking into account the dataflow heterogeneity of accelerators and memory bandwidths contentions. By doing so, service providers can employ the most efficient scheduling policy for user requests, optimizing Service-Level-Agreement (SLA) satisfaction rates and enhancing hardware utilization.
The application of RELMAS to a heterogeneous multi-accelerator system composed of various instances of Simba and Eyeriss sub-accelerators resulted in up to a 173\% improvement in SLA satisfaction rate compared to state-of-the-art scheduling techniques across different workload scenarios, with less than a 1.5\% energy overhead.
\end{abstract}

\maketitle

\section{Introduction}
Modern real-world applications present increasingly sophisticated challenges that demand equally complex solutions. Among various approaches, Deep Learning (DL), particularly Deep Neural Networks (DNNs), has demonstrated unparalleled capability in addressing these complex problems across diverse applications such as image recognition, object detection, natural language processing, IoT, social networks, medical diagnosis, drug discovery, autonomous driving, and service robotics. As a result, the demand for and reliance on DNNs have significantly grown in various domains.

However, executing DNNs is neither straightforward nor inexpensive due to extensive hardware requirements. With the increasing number of DL models developed daily, incorporating a rising number of parameters, greater memory, and computational resources become necessary~\cite{xu_nature18}. Although computational requirements continue to increase, the same cannot be said for computational power, \frar{partly }due to the slowdown of Moore's law and Dennard scaling. A common solution nowadays is to use Domain-Specific Accelerators (DSAs) for executing DNNs.

On the other hand, applications seeking fast, easy, and cost-effective solutions are increasingly delegating DNN execution to cloud services, which benefit from the computational power of large accelerators available in data centers. In the field of Engineering, there is a trend towards simplifying complex tasks, making them accessible to users and developers without extensive technical knowledge. This trend is evident in paradigms like FaaS (Function-as-a-Service, \ie Serverless Computing)~\cite{baldini_racc17}, which is now extended to Machine Learning tasks through the INFaaS (Inference-as-a-Service) paradigm~\cite{romero_usenix21}. Software developers are delegating such tasks to cloud services, enabling them to focus on core application aspects. As a result, efficient hardware resource management becomes crucial for service providers.

A promising near-term solution is to develop scheduling algorithms that efficiently map different neural networks from different tenants onto the same hardware resource, which typically is a multi-accelerator system (MAS), composed of cores called sub-accelerators (SAs). These algorithms must guarantee user-provided constraints, particularly ensuring that DNN model instances are executed within specified maximum deadlines.

This paper presents RELMAS, an approach to address the challenge of concurrent scheduling of various DNN models on a MAS with heterogeneous SAs. The main objective is to ensure diverse Quality of Service (QoS) levels, particularly in meeting maximum deadline constraints. RELMAS is an online scheduling method that creates a model of the environment, consisting of a MAS with a fixed number of SAs, and a varying number of instances of different DNN models that arrive online and need to be scheduled and completed within specified maximum deadlines. The proposed solution employs Deep Reinforcement Learning, specifically the Deep Deterministic Policy Gradient (DDPG) algorithm. We decided to unconventionally coordinate DDPG algorithm with Long-Short-Term-Memory (LSTM) networks, thus providing a technique to automatically understand deadline constraints among different DNN models instances and provide a schedule that effectively balances the workload through space and time. 
The algorithm effectively converges, resulting in a substantial reduction in deadline misses and a significant increase in deadline hits.

\ifdefined\longversion
The rest of the paper is organized as follows.
After an overview of existing contributions in Sec.~\ref{sec:rel_work}, Sec.~\ref{sec:prob} will introduce the specific problem addressed in this paper. Our proposed solution will be detailed in Sec.~\ref{sec:proposed_solution}, providing foundational terminology of reinforcement learning, and delving into the scheduling and learning phases. An evaluation will be conducted in Sec.~\ref{sec:eval}, and Sec.~\ref{sec:concl} will wrap up the paper by summarizing our contributions and highlighting potential avenues for short-term future research.
\fi


\section{Related Work} \label{sec:rel_work}
While the topic of multi-tenancy DNN scheduling in conventional general-purpose cores has been widely studied in both software and hardware since the introduction of chip-multi-processors, addressing multi-tenancy execution in DSAs is a more recent endeavor. Diverse approaches have emerged from academic research that can be classified into two categories~\cite{kim_hpca23}: static and dynamic. Static mechanisms involve configuring software or hardware statically (i.e., no decisions are taken at runtime) to adjust the memory access rates of contentious applications. Dynamic mechanisms, on the other hand, utilize runtime information to adaptively adjust the contentious nature of an application based on the actual amount of memory traffic in the system. 

Prior works like LayerWeaver~\cite{oh_hpca21}, PREMA~\cite{choi2020prema}, and AI-MT~\cite{baek_isca20} propose time-multiplexing DNN execution, either statically or dynamically. However, they often suffer from low hardware utilization, especially when layers cannot fully utilize all available resources. To enhance hardware utilization, static partition techniques such as HDA~\cite{kwon_hpca21} and MAGMA~\cite{kao_hpca22} propose spatial partitioning of compute or memory resources. However, their static nature hinders adaptability to different dynamic scenarios. While AI-MT employs multi-accelerators with homogeneous cores (specifically, systolic array architecture), MAGMA also supports multi-accelerators with heterogeneous cores. Additionally, the multi-tenancy support in AI-MT relies on heuristics, while MAGMA uses a research process based on a genetic algorithm. AI-MT considers potential inter-layer dependencies, unlike MAGMA. MAGMA is an offline scheduling strategy, requiring all job characteristics and arrival times to be known beforehand, which does not align with the requirements of on-demand services.
\frar{Recently, dynamic spatial partition mechanisms have been suggested to further improve co-location efficiency. }Planaria~\cite{ghodrati2020planaria} and Veltair~\cite{liu_asplos22} propose dynamic allocation of compute resources for co-running workloads. Despite their ability to adaptively allocate compute resources, such as processing elements for Planaria or CPU cores for Veltair, they lack the capability to dynamically manage memory resources.\frar{ This leads to low resource utilization and high thread migration overhead when repartitioning compute resources.} In contrast, MoCA~\cite{kim_hpca23} highlights the importance of memory-centric resource management, proposing a dynamic partition of both compute and memory resources throughout execution.

Although homogeneous-core accelerators have been thoroughly explored in the literature, heterogeneous architectures, often incorporating varied dataflows, enable more efficient processing of a diverse range of DNN layers, serving as both flexible and cost-effective solutions~\cite{symons2022towards}. However, in a data center environment that provides on-demand inference services -- with unpredictable job arrivals and Service Level Agreements primarily focused on execution within set deadlines -- the necessity arises for RELMAS, a sample-efficient, online DNN scheduling algorithm, that mitigates the risk of resource underutilization and QoS violations while maintaining bandwidth constraints.


\section{Problem formulation} \label{sec:prob}

\begin{figure}
    \centering
    \includegraphics{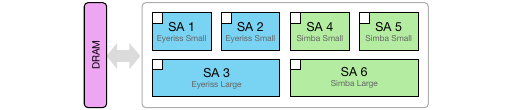}
    \caption{Diagram of the reference Multi-Accelerator Heterogeneous Architecture used in evaluations.}
    \label{fig:refarch}
\end{figure}


We consider a fixed hardware accelerator architecture, which consists of $M$ heterogeneous sub-accelerators (SAs), such as the one depicted in~Fig.\ref{fig:refarch}. In the context of this multi-accelerator system (MAS), individual SAs exhibit distinct specifications encompassing dataflow capabilities, processing unit counts, memory hierarchy configurations, and buffer sizes. All the SAs share the off-chip memory bandwidth. Such a system, which can ideally be deployed in both data center and edge scenarios, operates in a multi-tenant fashion, meaning it can receive and fulfill inference requests from multiple users. The resulting workload within the MAS can be parallelized at the layer granularity, allowing for concurrent processing of different layers on different SAs.

An user request comprises the task of executing a DNN model inference while adhering to latency constraints specified in a Service Level Agreement (SLA). We refer to each request as a job. Each job consists of several sub-jobs (SJs): one for each layer of the requested DNN model. At runtime, all the sub-jobs to be executed are collected in a virtually infinite ready queue (RQ). SJs are added to the RQ when a new request arrives and are removed when they start their execution on a SA or their deadline (stemming from the SLA) passes.

The objective of this work is to design a scheduling algorithm with the primary goal of maximizing the SLA satisfaction rate, namely the fraction of jobs completed while meeting the specified latency requirements, or conversely minimizing the deadline misses.

We make the reasonable assumption that all potential DNN models that may be requested are known in advance. This is in line with prior works~\cite{art:microsoft_soifer}. Layer dependencies and parameters of all these models are known. Thus, it is possible to compile a table containing: the latencies $c^m_{i,s}$ and the required bandwidths $b^m_{i,s}$ to execute the \mbox{$s$-th} SJ of the \mbox{$i$-th} potential request on the \mbox{$m$-th} SA. Due to the heterogeneity of the MAS, each layer executed on different SAs exhibits varying latency, energy, and bandwidth requirements. However, in an online context, during periods of low system activity, any unfamiliar model could undergo a ``registration phase'', involving compiling latency tables for the new model. Once registered, these models are integrated into the system. 

On the other hand, arrival times of each request $a_j$ are not known in advance. Thus, an online scheduling strategy is needed. Such a scheduler should determine the start time~$s_j$ for each SJ and the SA on which the SJ will be executed. A deadline miss will happen if the finish time $f_j > s_j + q_j$, where $q_j$ is the latency requirement of the request. The finish time cannot be simply calculated summing the computational costs of each layer on the assigned SA because of the limited memory bandwidth shared by all the SAs. When the sum of the bandwidth required by multiple SJs executing in parallel on multiple SAs exceeds the available memory bandwidth, all the SJs experience a slowdown proportional to the respective required bandwidth. This translates to the same amount of stall cycles for all the overlapping sub-jobs. Hence, it is important for the online scheduler to wisely orchestrate the execution of compute-intensive and memory-bound sub-jobs in the RQ to minimize bandwidth contentions.

The described problem can be considered as a generalization of the Flexible Job Shop Scheduling (FJSP) problem and, in Graham's notation~\cite{graham1979optimization}, it can be expressed as $FJSP\ |\ prec\ |\ U$, where $prec$ refers to the presence of precedence constraints among the SJs and $U$ is the number of tardy jobs, which is the objective to be minimized. It has been proved that the FJSP problem is NP-hard~\cite{dauzere2023flexible}. 

\section{RELMAS Online Scheduler} \label{sec:proposed_solution}

\begin{figure*}
     \centering
     \begin{subfigure}[b]{0.60\textwidth}
         \centering
         \includegraphics[width=\textwidth]{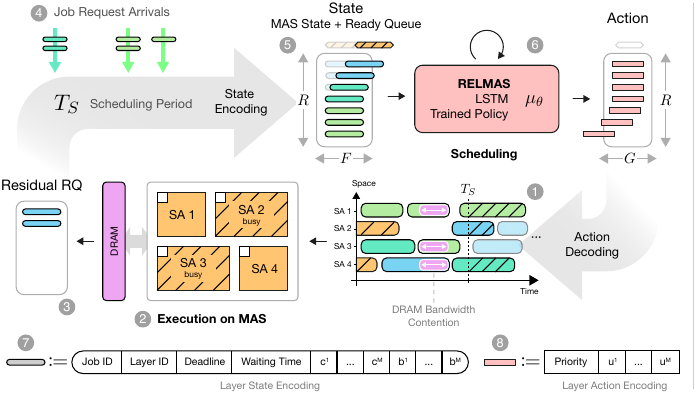}
         \caption{High-level visualization of the RELMAS policy in action during deployment.}
         \label{fig:scheduling}
     \end{subfigure}
     \begin{subfigure}[b]{0.30\textwidth}
         \centering
         \includegraphics[width=\textwidth]{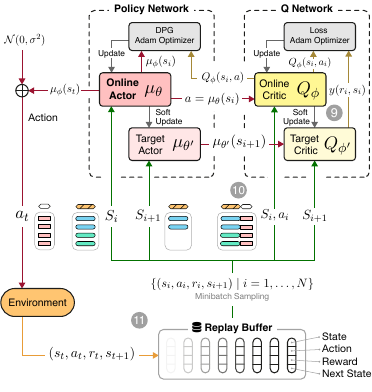}
         \caption{RELMAS policy learning process.}
         \label{fig:learning}
     \end{subfigure}
     \caption{Overview of the proposed online scheduler in production and learning phases.}
     \label{fig:overview}
\end{figure*}

\ifdefined\longversion
In this section we introduce RELMAS, the proposed approach to online scheduling in a multi-accelerator heterogeneous system based on Deep Reinforcement Learning (DRL). We adopt a top-down approach for this section: after an introduction about RL concepts, we start by depicting the working of RELMAS in a deployment setting, we then go deeper in the RELMAS agent learning process.
\else
In this section we introduce RELMAS, the proposed approach to online scheduling in a multi-accelerator heterogeneous system based on Deep Reinforcement Learning (DRL). For this section, we adopt a top-down approach: we first depict the working of RELMAS in a deployment setting, then we deepen the RELMAS agent learning process. For an introduction to the fundamentals of deep reinforcement learning, please refer to~\cite{arulkumaran_spm17}.
\fi

\ifdefined\longversion
\subsection{Reinforcement Learning Concepts}

To apply DRL effectively to any problem, some key concepts must be considered. First, there is the necessity of defining an environment, which serves as the context in which a learning agent operates. The environment encapsulates the dynamics and rules governing the system under study. Secondly, a state representation is crucial, as it provides a snapshot of the relevant information about the environment at a given time, enabling the agent to make informed decisions. Concurrently, an action encoding is required to specify the actions that the agent can take within the environment. Finally, the reward function plays a pivotal role, serving as the feedback mechanism that guides the agent's learning process. It provides a numerical measure of the agent's performance, with the goal of maximizing cumulative rewards over time.
\fi

\subsection{Scheduling}

In the target scenario, the environment consists in the MAS executing DNN layer inferences and receiving new requests. The latter represent the stochastic component of the environment. The state of the environment at any given time includes the current system load information, \ie the number of cycles for which each SA will be busy executing a previously assigned SJ, and the ready queue (RQ), \ie the collection of SJs ready to be executed. The action, which is the decision taken by the RELMAS policy agent for each SJ, can be articulated in a priority, defining temporal scheduling, and a SA choice, defining spatial allocation of the SJ in the MAS.
Fig.~\ref{fig:scheduling} shows an overview of RELMAS in a deployment setting. It can be noted that the scheduler is triggered periodically with a period $T_s$. During a period, the MAS execute \frar{a previously elaborated scheduling}{the latest elaborated schedule}, which is represented graphically in the space-time graph in Fig.~\ref{fig:overview}.1. At the end of the period, SJs remained unexecuted are collected in the so called residual RQ (Fig.~\ref{fig:overview}.3). These SJs are merged with the new SJs related to job requests arrived and queued in the meantime (Fig.~\ref{fig:overview}.4). 

This overall RQ (Fig.~\ref{fig:overview}.5) is sorted in ascending order of absolute deadline and submitted for scheduling by the RELMAS policy. We decided to implement the RL policy as a Long-Short-Term-Memory (LSTM) recurrent neural architecture. The rationale is that it helps dealing with a state of variable size (variable number of SJs in RQ) and LSTMs are good in detecting temporal relationships and patterns of SJs sequences. The LSTM cell with a hidden size of $h$ is succeeded by two fully connected layers: the first projects to $h/2$ units and applies a Rectified Linear Unit (ReLU) activation function, while the second projects to $G$ values and applies Hyperbolic Tangent (Tanh) activation function, encoding the decision, \ie the action, for the corresponding SJ.

As shown in Fig.~\ref{fig:overview}.6, the LSTM sequentially ingests the encoded representation of each SJ in the RQ. Hence, if $R$ sub-jobs reside in the RQ, $R$ timesteps of the LSTM policy are executed. The encoded representation (Fig.~\ref{fig:overview}.7) of each SJ is an array of features including all the information that could be useful to derive an optimal scheduling: an identifier of the model requested of which the SJ represent a layer, an identifier of the layer, the deadline representing the latency constraint\footnote{The deadline for each job request is explicitly stated and reiterated within the representation of each SJ. This redundancy enables the policy to differentiate between two identical SJs (same layer, same model), each associated with distinct request latency constraints.}, the waiting time informing about the distance in time from the arrival time of the request, the computational times and bandwidth requirements on the SAs. Thus, the length $F$ of each encoding is equal to $4+2M$.

In Fig.~\ref{fig:overview}.5 it is also possible to spot an extra element prepended to the encoded representation of the RQ. As no preemption is assumed, it is important to instruct the policy about the duration for which each SA will be busy, we do this by using a virtual SJ as a primer for the LSTM. All its features are zero except for computational times that instead represent the number of cycles before each SA will be available.

As output of the LSTM policy, after discarding the first timestep (corresponding to the primer) we get a decision encoding (Fig.~\ref{fig:overview}.8) for each SJ in the ready queue. Due to the Tanh activation function, all the output values relies in the $[-1,1]$ range that for each SJ are: the temporal priority and one value $u_m$ for each SA. The index associated with the highest $u_m$ determines the spatial allocation of the SJ in the MAS. Thus, the length $G$ of each encoding is equal to $1+M$.

\subsection{Learning}

As shown in Fig.~\ref{fig:learning}, we employ Deep Deterministic Policy Gradient (DDPG) which has emerged as a prominent algorithm in the field of reinforcement learning, particularly for solving high-dimensional continuous action space problems. For the sake of brevity, we will not describe all the mathematical details of DDPG, please refer to~\cite{lillicrap2015continuous} for a detailed explanation. In this section, we instead focus on the choices made to adapt DDPG to the problem addressed by this paper.

As in other RL algorithms, the goal in DDPG is to learn an optimal policy that maximizes a cumulative reward signal over time. To accomplish this, DDPG employs two essential components: the actor and the critic networks. Both are aimed to construct parameterized functions that iteratively converge to the corresponding optimal functions. Due to the exceptional capability of DNNs to approximate non-linear functions, two neural networks are used for these components.
The actor network represent the parameterized deterministic policy function $\mu_\theta$ and provides a continuous action output given an input state representation. On the other hand, the critic network estimates the Q-function $Q_\phi$, which quantifies the expected cumulative reward of following a given policy for a specific state and action pair. 

We already discussed the policy as an LSTM-based architecture. Likewise, we propose to adopt the same architecture for the critic network (Fig.~\ref{fig:overview}.9). The state and action pairs are concatenated horizontally as shown in Fig.~\ref{fig:overview}.10 and provided as input to the critic for Q-value estimation. Thus, the input length at each timestep is equal to $F+G$, while only one output value $Q_\phi(s,a)$ per timestep is projected from the last hidden state.

As in the baseline DDPG algorithm, experiences are stored in a \textit{replay buffer} (Fig.~\ref{fig:overview}.11) and sampled from it during the training phase for weight updates. Each experience is a tuple $(s_t,a_t,r_t,s_{t+1})$: $s_t$ is the environment state encoding at time $t$, $a_t$ is action representation of the decision taken by the policy, $r_t$ is the reward value obtained as an effect and $s_{t+1}$ is the state representation after executing the action. As we assume RELMAS to work in a periodic fashion and due to the stochastic request arrivals, $s_{t+1}$ can also include stochastic request arrivals, therefore breaking a deterministic causality chain from $s_{t}$ to $s_{t+1}$. Indeed, we found out that learning performance increases when $s_{t+1}$ only encodes the residual RQ instead of the overall RQ, eliminating the stochastic component from the experiences and allowing the algorithm to effectively infer causality chains.

Regarding the reward function, we defined it as follows:

\begin{align*}
    r_t &= \sum_{l \in {RQ}_t} \Delta_{t,l} \left( A_{t,l} + \gamma \frac{(a_l + q_l) - f_l}{q_l} \right) 
\end{align*}
\begin{multicols}{2}
\noindent\begin{equation*}
    \Delta_{t,l} = \begin{cases}
    1,  &  f_l < t + T_s \\
    \delta,  &  \text{otherwise}
\end{cases}
\end{equation*}
\noindent\begin{equation*}
    A_{t,l} = \begin{cases}
    \alpha, & f_l \leq a_l + q_l \\
    -\beta, & \text{otherwise}
\end{cases}
\end{equation*}
\end{multicols}

where $\delta \in [0,1]$ quantifies the relevance of events beyond the next scheduling period, $\alpha$ is the deadline hit reward, $\beta$ is the deadline miss penalty and $\gamma$ weights the significance of SJs normalized slacks. The introduction of normalized slacks, implemented as a scaled sum term, effectively mitigates resource underutilization. While compliance with SLA requirements is the foremost priority, this strategy additionally incentivizes the reduction of resource usage duration. This alignment serves the dual objective of not only fulfilling service commitments but also boosting operational efficiency.



\section{Evaluation} \label{sec:eval}
In this section, we present the evaluation of our novel online scheduling technique. To conduct our evaluation, we considered the hardware specifications detailed in Table~\ref{tab:archspec}. Our chosen hardware configuration is rooted in a heterogeneous accelerator architecture following the principles described in~\cite{kwon_hpca21}, boasting an equivalent off-chip memory bandwidth and total number of MAC units as in other works focusing on systolic arrays~\cite{kim_hpca23}. We are aiming for a multi-chip-module-based architecture in which sub-accelerators take the form of chiplets interconnected by means of a network-on-package (NoP). The on-package links utilize a silicon interposer and incorporate efficient intra-package signaling circuits, as in~\cite{shao_micro19}. The selected architectural configuration comprises six sub-accelerators, which are instances of two widely recognized state-of-the-art DNN accelerators, as visualized in Fig.\ref{fig:refarch}. Specifically, half of these sub-accelerators adopt a row-stationary dataflow and an internal organization closely resembling Eyeriss\cite{chen2019eyeriss}, while the remaining half employ a weight-stationary dataflow and an architecture akin to a Simba~\cite{shao_micro19} chiplet. We have chosen to integrate both small and large instances from these two categories of sub-accelerators (see Table~\ref{tab:archspec}). This strategic decision enables more effective orchestration of memory-bound and compute-intensive workloads.
\begin{table}
  \caption{Parameters of the sub-accelerators classes considered in evaluations.}
  \label{tab:archspec}
  \begin{center}
  \footnotesize
  \begin{tabular}{l cccc}
    \toprule
     & \multicolumn{2}{c}{\textbf{Eyeriss}} & \multicolumn{2}{c}{\textbf{Simba}} \\
    \cmidrule(){2-3}\cmidrule(l){4-5}
    \textbf{Parameter} & \textbf{Small} & \textbf{Large} & \textbf{Small} & \textbf{Large} \\
    \midrule
    Frequency &
    \multicolumn{4}{c}{1 GHz} \\
    DRAM Bandwidth &
    \multicolumn{4}{c}{16 GB/s} \\
    NoP Bandwidth, Energy &
    \multicolumn{4}{c}{100 GB/s, 1.3 pJ/bit } \\
    Dataflow & \multicolumn{2}{c}{Row Stationary} & \multicolumn{2}{c}{Weight Stationary} \\
    Num. PEs & 256 & 512 & 16 & 32 \\
    MACs per PE & 1 & 1 & 16 & 16 \\
    Global Buffer & 64 KiB & 64 KiB & 32 KiB & 64 KiB \\
    PE Buffers & 220 B & 220 B & 24 KiB & 24 KiB \\
    \bottomrule
  \end{tabular}
  \end{center}
\end{table}

\begin{table}
  \caption{Benchmark workloads used in evaluations.}
  \label{tab:workloads}
  \begin{center}
  \footnotesize
  \begin{tabular}{ccc}
    \toprule
    \textbf{Workload} & \textbf{Domain} & \textbf{DNN Models} \\
    \midrule
    Light & Image Classification & SqueezeNet \\
    & Object Detection & YOLO-Lite \\
    & NLP & Keyword Spotting \\
    \midrule
    Heavy & Image Classification & AlexNet, InceptionV3, ResNet50 \\
    & Object Detection & YOLO-V2 \\
    \midrule
    Mixed & \multicolumn{2}{c}{Light $\cup$ Heavy} \\
    \bottomrule
  \end{tabular}
  \end{center}
\end{table}
In our evaluations we consider seven different state-of-the-art DNN models from image classification, object detection and natural language processing (NLP) domains. 
These DNN models features different model sizes, computational, memory and bandwidths requirements rendering them highly representative of a wide spectrum of DNN workloads. Similarly to~\cite{ghodrati2020planaria} and~\cite{kim_hpca23}, we organized these models in three workload sets classified by size as summarized in Table~\ref{tab:workloads}. To generate multi-tenant scenarios from each of these three workload sets, we consider a time interval during which inter-arrival times of inference requests are drawn from a Pareto distribution, emulating task dispatching in data centers according to insights from~\cite{art:googleworkload}. 
\ifdefined\longpaper
For each request the specific DNN model is selected randomly from the workload set under consideration. As already described, each request is a job, consisting of as many sub-jobs as layers in the DNN model.
\fi

Each request is associated with a SLA requirement defined by the QoS level. Adopting the PREMA \cite{choi2020prema} approach, the latency requirement is calculated multiplying a coefficient, that we name QoS factor, by the minimum execution latency of the requested job (\ie without interference of other jobs). Specifically, we identify a QoS factor for the baseline QoS level (\mbox{QoS-Medium}), then we derive other two targets: \mbox{QoS-Low} which indicates relaxed latency constraints corresponding to \mbox{$1.2\times$ QoS-Medium} and QoS-High denoting \mbox{$0.8\times$ QoS-Medium}, similarly to~\cite{ghodrati2020planaria} and~\cite{kim_hpca23}.

We assess the energy, latency and memory bandwidth requirements figures of each DNN layers on each sub-accelerator using the validated Timeloop/Accelergy~\cite{parashar2019timeloop} simulation infrastructure. Additionally, we develop a multi-accelerator multi-tenant simulation platform that takes into account layer latencies, layer dependencies and runtime memory bandwidth contentions to accurately determine the start and finish times of each sub-job when executing according to a specific scheduling algorithm. This methodology allows us to effectively evaluate the target metrics. Across all the evaluations, the following reward coefficients were used during RELMAS learning: $\alpha=0.10$, $\beta=0.11$, $\gamma=0.05$ and $\delta=0.01$.

\subsection{SLA Satisfaction Rate}

\begin{figure*}
     \centering
     \begin{subfigure}[b]{0.28\textwidth}
         \centering
         \includegraphics[width=\textwidth]{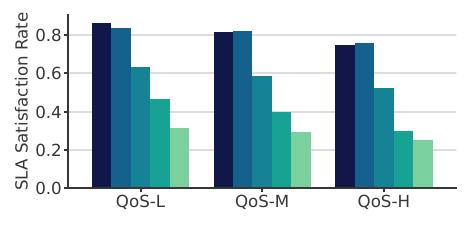}
         \caption{Light Workload}
     \end{subfigure}
     \hfill
     \begin{subfigure}[b]{0.28\textwidth}
         \centering
         \includegraphics[width=\textwidth]{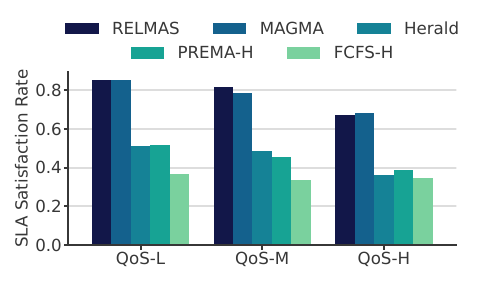}
         \caption{Heavy Workload}
     \end{subfigure}
     \hfill
     \begin{subfigure}[b]{0.28\textwidth}
         \centering
         \includegraphics[width=\textwidth]{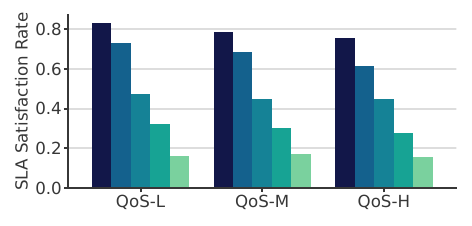}
         \caption{Mixed Workload}
     \end{subfigure}
        \caption{SLA Satisfaction Rate comparison against other baselines for different workload sets.}
        \label{fig:slart}
\end{figure*}

In Fig.~\ref{fig:slart} we show the results of the evaluation of the three workload sets in Table~\ref{tab:workloads} with three different QoS targets. We adopt the SLA Satisfaction Rate as our primary comparative metric.
To evaluate the efficacy of the proposed approach, we conduct comparative analyses against robust baselines, which draw inspiration from and are adapted from previous works in the field. The following baseline algorithms are considered: (1) FCFS-H, which is the First-Come-First-Served scheduling algorithm for priority assignment to each sub-job in the ready queue, paired with an heuristic for the selection of the sub-accelerator. The heuristic consists in choosing the SA which minimizes the finish time of the sub-job. (2) PREMA-H, which is a scheduling algorithm based on \cite{choi2020prema}, combining a token mechanism based on waiting time and Shortest-Job-First priority assignment. As the original PREMA work targets a monolithic architecture, for a fair comparison we combine it with the aforementioned heuristic. (3) Herald \cite{kwon_hpca21}, which is a layer scheduling algorithm designed for heterogenous architectures, focusing on load balancing of the SAs. (4) MAGMA \cite{kao_hpca22}, consisting in a genetic algorithm approach to scheduling optimization. Our implementation adopts the same custom genetic operators of the original work. For a fair comparison, we used the same evaluation platform as RELMAS and adapted its fitness function to take into account the SLA satisfaction rate.
\ifdefined\previousVersionWithItemize
\begin{itemize}
    \item FCFS-H, which is the First-Come-First-Served scheduling algorithm for priority assignment to each sub-job in the ready queue, paired with an heuristic for the selection of the sub-accelerator. The heuristic consists in choosing the SA which minimize the finish time of the sub-job.
    \item PREMA-H, which is a scheduling algorithm based on \cite{choi2020prema}, combining a token mechanism based on waiting time and Shortest-Job-First priority assignment. As the original PREMA work targets a monolithic architecture, for a fair comparison we combine it with the aforementioned heuristic.
    \item Herald \cite{kwon_hpca21}, which is a layer scheduling algorithm designed for heterogenous architectures, focusing on load balancing of the SAs.
    \item MAGMA \cite{kao_hpca22}, consisting in a genetic algorithm approach to scheduling optimization. Our implementation adopts the same custom genetic operators of the original work. For a fair comparison, we used the same evaluation platform as RELMAS and adapted its fitness function to take into account the SLA satisfaction rate.
\end{itemize}
\fi

All the evaluated algorithms are supposed to be triggered for each scheduling period at runtime. For each tested scenario we trained RELMAS considering a hidden size of 256 for both the actor and the critic. As shown in Fig.~\ref{fig:slart}, the trained policy is able to match or outperform the other methods across all the scenarios in terms of SLA satisfaction rate. 

With respect to Herald, RELMAS achieves a geometric mean improvement of $59.4\%$, while with respect to \mbox{PREMA-H} it achieves a geometric mean improvement of $109\%$. The gratest improvements with respect to these two baselines are achieved for the Mixed workload. We believe, this effect should be attributed to high variability of model sizes in the Mixed workload, that cause the scheduling process to be more complex. In particular, the Shortest-Job-First strategy of PREMA, allocates more priority to the Light models in the Mixed workload starving the Heavy models.

RELMAS achieves the same SLA Satisfaction Rate of MAGMA in the Light and Heavy workloads. However, it is worth noting that MAGMA is executed with same parameters of the original paper, \ie 100 generations of 100 individuals, for each scheduling period to optimally scheduling a ready queue. For the evaluation of individual fitness, we employ the identical simulation platform, previously delineated and utilized for RELMAS training. These choices resulted in MAGMA scheduling runtimes of several minutes on a standard workstation for each simulated scheduling period, which is reasonable as MAGMA, unlike RELMAS, was born to be an offline scheduling policy, not an online one. While this places our MAGMA implementation in the realm of limited practicality within real-world contexts, it serves as a strong benchmark for evaluating the proposed approach. Furthermore, RELMAS is capable to outperform MAGMA in a Mixed workload workload scenario with up to $22.5\%$. SLA Satisfaction Rate improvement. We will analyze the overhead of the proposed solution in Sec.~\ref{sec:overhead}.

\subsection{Bandwidth Sensitivity}

\begin{figure}
    \includegraphics[width=0.88\columnwidth]{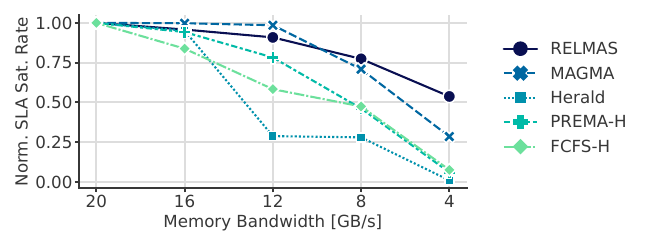}
    \caption{Impact of memory bandwidth reduction on SLA Satisfaction Rate for different scheduling strategies.}
    \label{fig:bandwidth}
\end{figure}

RELMAS is designed to be aware of the bandwidth required by each DNN layer to be executed without stalls. To this end, this information is embedded as a feature in encoded LSTM policy input. This makes the proposed technique more robust in memory-constrained contexts. We test a Light workload scenario and in Fig.~\ref{fig:bandwidth} we show the SLA Satisfaction Rate of different scheduling strategies normalized to their respective best result achieved. It can be noted that, independently from their absolute performance, some scheduling approaches tend to experience an abrupt degradation when decreasing the memory bandwidth, with respect to their performance in a bottleneck-free situation. On the other hand, RELMAS degradation is slower, especially in the low-bandwidth region, indicating an higher capability of the proposed approach to better orchestrate layers on the heterogeneous hardware optimizing memory bandwidth utilization. We found that the same conclusions remain valid for the other considered workloads; therefore, we are reporting the results for the Light workload only.


\subsection{Overhead Assessment}

\begin{figure}
    \centering
    \includegraphics[width=0.88\columnwidth]{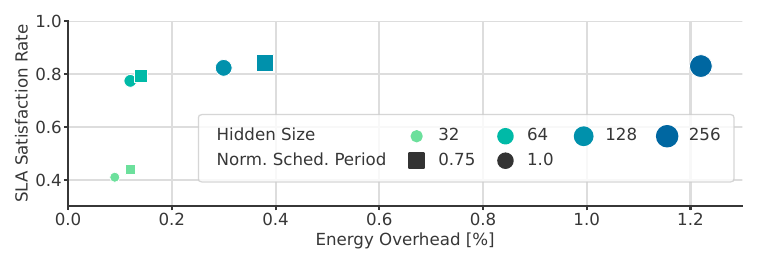}
    \caption{Energy overhead of the proposed scheduling algorithm varying the hidden size of the LSTM policy and the scheduling period. }
    \label{fig:overhead}
\end{figure}

We now analyze the overhead of the proposed solution for the online scheduling in a multi-accelerator system. As already described, RELMAS consists in a DNN model composed of an LSTM cell followed by two projection fully connected layers. The policy works recurrently ingesting one input state encoding at each timestep, one for each layer in the ready queue. To gain insight into the computational complexity, it is informative to note that for a policy with an hidden size of 256, the number of Multiply-Accumulate (MAC) operations per layer amounts to \numprint{316288}, representing about a mere $0.04\%$ and $0.007\%$ of the MAC operations required by the AlexNet and ResNet50 models respectively. 
We opted to implement the scheduling policy within one of the accelerators designated for the policy’s execution. This was achieved by communicating the accelerator's busy status to the scheduler, which was done by adding the overhead time to the first element in the State encoding, as shown in Fig.~\ref{fig:overview}.5. Our experiments revealed that this implementation incurred no overhead impact on the scheduling process itself.
We conducted an accurate evaluation of the energy overhead assuming that the scheduler is executed on the MAS itself. In particular, we assumed that one of the Simba Small SAs (Fig.~\ref{fig:refarch}) is responsible for running each timestep of the RELMAS policy. We evaluated the energy required for this setup using Timeloop~\cite{parashar2019timeloop}. We then assessed the overall energy overhead to the workload execution considering different hidden sizes for the LSTM cell and the Mixed workload. The results are shown in Fig.~\ref{fig:overhead}. We found that, for the considered workload, there is no significant SLA Satisfaction Rate improvement for an hidden size bigger than 128. In any case, the energy overhead does not exceed $1.3\%$. We also noticed that decreasing the scheduling period can increase the performance of the scheduler, because it enables shorter reaction times. This comes to the cost of a slight increase in energy overhead due to the increasing average length of residual ready queues. In other words, a layer will be scheduled multiple times on average before being actually executed, because it is often forwarded to the next scheduling period.
\label{sec:overhead}

\section{Conclusion} \label{sec:concl}
This paper presents RELMAS, a low-overhead deep reinforcement learning based online scheduling policy for DNN multi-tenant heterogeneous multi-accelerator systems.
We believe that reinforcement learning could be a viable and low-overhead solution for online scheduling and dispatching in multi-accelerator multi-tenant heterogenous systems. For the proposed methodology, we included computation times and memory bandwidths requirements as features for the RL policy. We think many more could be explored to enable other functionalities. 
For instance, tenant-wise SLA satisfaction and fairness awareness could be enabled by exploring new state encoding and online tenant-wise deadline misses closed-loop control with dynamic reward shaping. Furthermore, we unconventionally considered an LSTM architecture for the policy but other trending architectures could be explored and their overhead assessed such as transformer-based models.

\begin{acks}
This work has been (partially) supported by MUR project ARS01\_00592 reCITY. The work of M.~Palesi, who has contributed to the development of Secs.~3 and~4, has been supported by the Spoke 1 "FutureHPC \& BigData" of the Italian Research Center on High-Performance Computing, Big Data and Quantum Computing (ICSC). The work of V.~Catania, who has contributed to Sec.~5, has been supported by PNRR MUR project PE0000013-FAIR.
\end{acks}

\bibliographystyle{abbrv}
\bibliography{bibliography}


\end{document}